# Exciton-impurity luminescence
# from noncrystalline xenon-argon clusters

**O.G. Danylchenko, Yu.S. Doronin, S.I. Kovalenko,**
**M.Yu. Libin, V.N. Samovarov, V.L. Vakula**

*B. Verkin Institute for Low Temperature Physics and Engineering*
*of the National Academy of Sciences of Ukraine*
*47 Lenin Ave., Kharkiv, 61103, Ukraine*
*E-mail: samovarov@ilt.kharkov.ua*

For the first time in binary mixtures of solid rare gases exciton-impurity luminescence is observed from a xenon-argon system containing argon as impurity. An exciton-impurity emission band is registered for binary clusters with the structure of multilayer icosahedron. The optical transition occurs from an energy level lying very close to the lowest level of volume excitons in bulk xenon samples. The results demonstrate the possibility of probing excitonic levels in noncrystalline condensed media.

**1.** In our previous works [1,2], we reported the first observation of luminescence from bulk and surface excitons in substrate-free xenon clusters. Earlier, excitons in pure and binary rare gas solid clusters were only observed in absorption spectra [3], which are formed over $10^{-14}$ s, however the dynamics of excitons during their life span of $10^{-9}$ s was unknown since no exciton luminescence spectra had been registered.

In Refs. [1,2] two situations were studied: (i) Xe clusters without Ar shell; (ii) Xe clusters covered with a thin Ar shell. A strong suppression of the exciton luminescence bands was observed in the second case, while in the first case the bands were clearly seen. It was found that bulk excitons, which have the diffusion length of up to 1000 Å in bulk cryocrystals, do not decay in small clusters of about 50 Å in diameter, i.e. when efficient channels exist through which they should decay upon interaction with the cluster surface [1].

From the other hand, it was shown in Refs. [4,5] that mixed Xe-Ar clusters undergo a phase separation at which Xe atoms form the cluster core, while the surface layers become enriched with the surface-active Ar admixture. Moreover, a combined electron diffraction and spectroscopic study [6] demonstrated that under some conditions there is a complete decay of the system into virtually pure solid components. Thus it seems interesting to study the evolution of excitons upon the decay of the primary Xe-Ar solution. This task is rather close to the long-standing problems of the physics of excitons in doped semiconductors where exciton-impurity complexes with great oscillator strengths and rearrangement of the exciton spectrum near the surface enriched with defects or impurity are observed [7]. As far as we know, there have been no studies undertaken so far to resolve the task set either with mixed clusters, or with bulk solid rare gas samples having impurities[1]. For nanosystems, such studies may be of special interest from the viewpoint of exciton spectroscopy of quantum dots with impurity.

Here we report the first observation of the change in intensity and energy position of the exciton bands upon decay of a Xe-Ar solution into pure components. It was rather surprising to observe exciton (polariton) emission bands from rather small Xe-Ar clusters containing about 1000 atoms ($R_{cl} \approx 25$ Å) and possessing the structure of multilayer icosahedron.

The most general conclusion that we can draw from the results of this work is as follows: luminescence of exciton-impurity complexes, which are a bound state of a Xe-matrix exciton and Ar impurity, has been observed for the first time from a noncrystalline medium, such as Xe-Ar icosahedral clusters.

**2.** In this work spectroscopic measurements were performed along with electron diffraction studies to obtain more reliable information about the cluster structure. Cathodoluminescence spectra were excited by a 1-keV electron beam and registered in the spectral range of 6-8.5 eV. Electron diffraction measurements yielded diffraction patterns from clusters for diffraction vectors up to $s \approx 6.5$ Å$^{-1}$. Clusters were obtained by condensation of an Ar-Xe mixture in a supersonic jet flowing into vacuum through a conical nozzle with the critical cross section 0.34 mm and cone opening angle 8.6°. Dimension and structure of clusters could be varied by varying Xe content $C_{Xe}$ in the primary gas mixture, as well as the mixture pressure $p_0$ and temperature $T_0$ at the nozzle entrance. In contrast to our previous papers [1,2], here we studied in detail Xe-Ar clusters with no Ar shell as Ar content in clusters was gradually increased. For this purpose we varied Xe concentration $C_{Xe}$ from 1.8 to 4.7% in the primary gas mixture kept at a constant pressure ($p_0=1$ atm) and temperature ($T_0=190-200$ K). For a more detailed description of the experimental method used see Refs. [4,8].

---

[1] In mixed Xe-Ar bulk samples with a limited solubility, the complete decay into pure components is not observed, there is instead a decay into solutions with a small content of the impurity component.

**3.** Figure 1 shows diffraction patterns obtained for Xe-Ar mixtures with the Xe concentrations $C_{Xe} = 2.5\%$ and 4.7% (*a*), and the cathodoluminescence spectrum for $C_{Xe} = 2.5\%$ (*b*). Diffraction maxima from solid xenon (111), (200), (220), (311) can be clearly seen, as well as the overlapping maxima (331) and (420). The (220) and (311) peaks are poorly discernible (which is not the case for pure Xe clusters [9]), indicating that there is a halo in that region resulting from Ar impurity atoms chaotically distributed inside the cluster.

The character of the diffraction patterns clearly suggests that we deal with a Xe multilayer icosahedron. This is evidenced by a relatively small intensity of the peaks found in the positions of the *hcc* maxima (220) and (311) with respect to the (111) peak intensity and by the maximum located in the position of the (200) peak and very much overlapped with the (111) maximum, see Fig. 1(*a*). The energy position of the (111) peak made it possible to estimate the weighted mean interatomic distance $a = 5.9$ Å in the clusters (both for $C_{Xe} = 2.5\%$ and $C_{Xe} = 4.7\%$), which is on the average 4.5% smaller than the lattice constant in the *hcc* structure of bulk Xe: $a=6.15$ Å ($T=40$ K) and $a=6.19$ Å ($T=70$ K). The obtained value confirms the multilayer ichosahedron structure of our clusters which is characterized by smaller distances between the internal (111) planes with respect to those of the crystalline *hcc* lattice [10]. For example, in a multilayer icosahedron having 6 Mackay spheres in total (atom number being 923), the distance between the 1st and 2nd spheres is 6.5% smaller, and between the 4th and 5th spheres is 3% smaller than in the *hcc* lattice[1].

As can be seen from Fig. 1, the $C_{Xe} = 2.5\%$ diffraction pattern does not reveal any maxima in the region of $s = 3.9$ Å$^{-1}$ (311); 5.3 Å$^{-1}$ (331+421) and 6.1 Å$^{-1}$ (333+511), which are typical of bulk argon. This suggests that Ar atoms do not form any several-monolayer shell on the surface of the Xe core. However, for $C_{Xe} = 4.7\%$ we have some blurred maxima at the position of the Ar peaks (331)/(420) at 5.3 Å$^{-1}$ and (333)/(511) at 6.1 Å$^{-1}$. This allows us to suppose that the noncrystalline component of argon increases significantly within the layers lying near the surface of the Xe cluster at the initial stage of phase segregation. It should be noted that the process leaves virtually unchanged the interatomic distance derived from the position of the Xe maxima since the Ar content is very much constant within the more deeply lying layers.

According to our electron diffraction data and the results of our previous studies (see, for example, Refs. [2,9]), the number of atoms in a cluster can be estimated to be about 1000; the cluster temperature, following Ref. [10], must be near 50 K.

Consider now the luminescence spectrum in Fig. 1*b*. It is multicomponent, the thin lines are Gaussian fit to the experimental curve and its components. Our previous studies (see Refs [1,2] and references therein) allow us to make the following assignments of the bands. The band at 8.353 eV (FWHM is 21 meV) is located closely to the position of the emission band of bulk $\Gamma(3/2)$ exciton with the quantum number n=1 seen from crystalline xenon samples (8.333 eV at $T$=50 K, FWHM being ≈9 meV [1]). The bands at 8.225 eV (35 meV) and 8.315 eV (21 meV) arise from the crystal field splitting of the surface exciton emission band, while the band at 8.38 eV was earlier observed in luminescence spectra of bulk samples of solid xenon and was assigned to a surface polariton[2].

---

[1] In bulk samples, no more than 5-7% of argon can be solved in xenon. Dissolved argon can cause contraction of the lattice, but to a much lesser extent than in the case of interplanar spacing diminution in a multilayer icosahedron [11].

The observation of exciton (polariton) luminescence bands from small xenon clusters having the structure of a multilayer icosahedron is rather peculiar. Some time ago, the registration of exciton emission bands in bulk crystalline rare gas samples [13] required development of special techniques for obtaining perfect crystals and their cleaning from atmosphere impurities.

The other luminescence bands can be assigned to the following transitions (see Ref. [2] and references in the discussion paragraph). The strong sharp line at 8.435 eV (11 meV) is due to the emission $^3P_1 \to {}^1S_0$ of a desorbed Xe atom away from the cluster surface. The energy position of the 8.295 eV band (26 meV) is rather close to the forbidden transition $^3P_2 \to {}^1S_0$, which can become allowed near the sample surface. The 8.266 eV band (35 meV) is due to the emission of the exciplex molecule (ArXe)$^*$ and was earlier observed upon bombardment of Ar cryocrystals by excited Xe atoms [14], which can provide additional evidence for the presence of Ar in the cluster.

Consider now the behavior of intensity and energy position of the luminescence bands from bulk and surface excitons upon variation of Xe concentration $C_{Xe}$ in the primary gas mixture. Fig. 2 shows the concentration dependence of integral intensities of the exciplex molecule (ArXe)$^*$, bulk and surface exciton bands. The exciplex (ArXe)$^*$ band intensity grows continuously with Xe concentration within the whole range of $C_{Xe}$. However, the intensities of the exciton bands are observed to grow for Xe concentrations increasing from 1.8 to 2.5% and to fall for higher concentration values ($C_{Xe} \geq$ 2.5-3.0%) to become several times smaller than their maximum value at $C_{Xe}$=4.7%. It should be noted that it is with Xe concentration increasing from 2.5 to 4.7% that we observed growth of the (Xe$_2$)$^*$ molecule luminescence band at 7.0 eV, which suggests, as in the case of bulk cryocrystals, localization of electrons in a disordered lattice accompanied by formation of a strongly bound excited molecular center (in an ordered lattice, this luminescence band is centered at 7.2 eV) [15].

Figure 3 shows the red shift of the 8.35 eV bulk exciton band upon an increase in Xe concentration in the primary gas mixture. There is no shift observed for the other bands (within the experimental errors). As can be seen, the shift remains small (being about 2 meV) for Xe concentrations growing up to ≈3%, but it becomes greater with further growth of Xe concentration.

---

[1] Due to the additional contraction of lattice in clusters, the matrix exciton levels should lie a bit higher than those in bulk cryocrystals (at the same temperature).
[2] It is known that solid xenon is characterized by a rather strong light-exciton coupling [12], i.e. exciton states in it can be spoken of as polaritonic. For the purposes of this paper, the differences between the terms 'exciton' and 'polariton' are not crucial, so in accordance with the established tradition for the solid xenon we will use both while describing the spectra below.

**4.** Before we begin to discuss the results, we'd like to note once more that for the purposes of this work Xe clusters were created with Ar atoms inside that did not form any Ar shell in contrast to the experiments performed in our Refs. [1,2]. The absence of Ar shell was confirmed by electron diffraction measurements, an independent evidence was also provided from the measured luminescence spectra. It was shown for binary Xe-Ar mixtures [1,2,16] that the presence of an outer shell made up of the lighter component atoms should be evidenced by the 8.41 eV band shifted to the redder end from the $^3P_1$ line at 8.44 eV that comes from desorbed Xe atoms. This band appears when an excited Xe atom is stabilized outside the cluster near its surface in the 'outer' minimum of the potential arising from its interaction with the Ar shell of the cluster. This bound state is only possible for heteronuclear interactions of an excited Xe (Kr) atom with an Ar shell possessing a negative electron affinity [16]. The absence of the band in our spectra (see, e. g., Fig. 1*b*) for all the initial Xe concentrations used suggests that Ar atoms are distributed inside the cluster without forming any Ar shell on its surface. However, under conditions close to those leading to the complete phase segregation the distribution of argon along the radial direction is not uniform.

Consider now the spectral shift of the bulk exciton luminescence band which is only observed upon variation of Xe concentration in the primary gas mixture. It has already been mentioned that, according to the diffraction data in Fig. 1*a*, an increase in $C_{Xe}$ is responsible for growth of Ar content in the cluster. Since the diffraction patterns don't reveal any noticeable change in interatomic distances, the observed shift of the exciton band should be attributed to a change in concentration of *argon* atoms in the cluster.

It is known that the red shift of exciton bands may also be caused by the formation of exciton-impurity complexes [7]. In our situation, this would correspond to the formation of a coupled state of a Xe matrix exciton and one or several Ar atoms (aggregates). The shift magnitude depends on coupling energy and is, according to our data, several meV, being different for different structures of the impurity center.

As the Xe cluster grows (upon increase in Xe concentration in the primary mixture), nanodroplets emerging at the first stage of the formation of solid aggregates become 'hotter' due to the release of latent heat of condensation. This facilitates the process of diffusion of surface-active atoms of argon into outer layers (i.e. the outer layers become excessively enriched with argon atoms) which is known to precede the radial phase segregation. Such enrichment increases the probability of formation of exciton-impurity complexes with Ar aggregates containing two, three, and more atoms.

Thus, the observed enhancement of the red shift in the concentration range over 2.5-3% can be accounted for by the assumption that as the number of Ar atoms increases in the cluster, not only exciton-impurity complexes with *one* Ar atom involved arise with a higher probability (especially near the cluster surface), but also those formed on Ar aggregates (possessing a greater coupling energy). At the same tine, the lattice gets deformed in the vicinity of such complexes and the localization barrier decreases, i.e. the probability of electron relaxation into more deeply lying levels of the molecular centers $(Xe_2)^*$ (revealing itself as a wide band centered at 7.0 eV) becomes higher. According to our data, the localization grows more intense for Xe concentrations in the primary gas mixture over 2.5-3%, i.e. when there is a strong shift of the exciton bands accompanied by a decrease in their intensities (see Fig. 2). The exciplex $(XeAr)^*$ band intensity increases in this concentration range, therefore the exciton localization results in formation of both excimer molecules $(Xe_2)^*$ and heteronuclear excited centers $(XeAr)^*$.

**5.** A general picture of the formation of exciton luminescence in a noncrystalline medium, such as icosahedron, can be presented as follows. In case there is no impurity, a newly created exciton should occupy deeply lying states of the excited molecular centers $(Xe_2)^*$ over the characteristic phonon times (due to the smallness of the energy barrier in a disordered medium). By localizing on the surface it also can cause desorption of excited atoms (of Xe in our case). However, if we have impurity centers (of argon) that cannot be ionized by excitons, the situation changes drastically: a new channel appears leading to the formation of exciton-impurity complexes. For small coupling energies, much smaller than the exciton band width, the exciton-impurity energy level lies very close to the lowest state of the matrix bulk exciton. We'd like to note that in rare gas solids the energy levels of bulk excitons are known to exist in clusters that contain 50-100 atoms and more, that is including those with the structure of multilayer icosahedron, which is confirmed by the measured luminescence excitation spectra (absorption spectra with the characteristic time scale of $10^{-14}$s) [3]. Moreover, according to optical absorption data an exciton energy level exists even in liquid xenon, when a short-lived cluster of an enhanced density of atoms arises in the fluctuating medium. However, over the time scale of 1-3 ps the exciton transforms into an excimer molecule $(Xe_2)^*$ [17].

The width of the lowest exciton band $\Gamma(3/2)$ in Xe is $\Delta \approx 1$ eV, the energy of the van der Waals coupling between the exciton and an Ar atom is $\varepsilon \approx 2$ meV, the interaction radius is about 6 Å, being close to the interplanar distance in solid xenon (calculations can be made by using the estimations from Ref. [18] and taking into account that the exciton radius is $\approx 3$ Å in Xe). It is known that in molecular and semiconductor crystals the ratio $\Delta/\varepsilon \approx 10^3$ is responsible for the giant enhancement of optical transition oscillator strength $f \propto (\Delta/\varepsilon)^{3/2}$ of an exciton-impurity complex [7]. We believe that it is this effect that makes it possible to observe the pronounced exciton features in emission spectra from a noncrystalline medium. As the content of argon increases in the cluster, interaction of excitons with Ar aggregates sets in. Since the coupling energy of such exciton-impurity complexes grows larger, a red shift is observed for the luminescence band. The maxima of the surface exciton and polariton emission bands turn out to be less sensitive to Ar content.

In case of the complete phase segregation, when the Xe core is covered with an Ar shell, almost all bulk excitons transform into deeply lying molecular-type local centers. The surface excitons of xenon are less sensitive to the Ar shell, while excited atoms of xenon possessing an excessive kynetic energy are even capable of penetrating the shell to emit radiation as desorbed atoms or those stabilized near the cluster surface [1,2,16].

The paper reports the first observation of exciton-impurity luminescence in binary solid rare gases registered in a Xe-Ar system, argon being an impurity. The relevant optical transition occurs from an energy level lying very close to the lowest state of Xe matrix bulk exciton. An intense exciton-impurity emission band is observed in binary clusters having the structure of multilayer icosahedron. It is quite possible to expect the emission band to appear in bulk Xe sample with an Ar impurity, including the fine-grained phase and even a glassy state that can emerge both in bulk samples and in rare gas clusters [19]. The results obtained provide new possibilities to study exciton levels in disordered solid rare gas media by generating exciton-impurity states with huge oscillator strengths.

**Figure captions**

**Fig. 1.** Diffraction patterns and luminescence spectrum from Xe-Ar clusters with a structure of multilayer icosahedron, obtained by condensation of a binary Ar-Xe gas mixture ($p_0 = 1$ atm, $T_0 \approx 200$ K, $C_{Xe} = 2.5\%$ and 4.7%). Diffraction patterns ((111) peak intensity taken as unity): solid line – $C_{Xe} = 2.5\%$, dashed line – $C_{Xe} = 4.7\%$, arrows indicate diffraction maxima from solid xenon and halo position from solid argon (*a*). Cathodoluminescence spectrum from Xe-Ar clusters: $C_{Xe} = 2.5\%$. Thin lines show spectrum fit by Gaussian components. Bulk exciton band at $\approx 8.353$ eV is shaded grey. Assignment of the other components to emission centers is given in the text (*b*).

**Fig. 2.** Integral intensity of (ArXe)* exciplex molecule and of bulk (8.35 eV) and surface (8.31 eVB) exciton emission bands as a function of Xe concentration in the primary Ar-Xe gas mixture.

**Fig. 3.** Energy shift of bulk exciton (exciton-impurity complex) emission band as a function of Xe concentration in the primary Ar-Xe gas mixture.

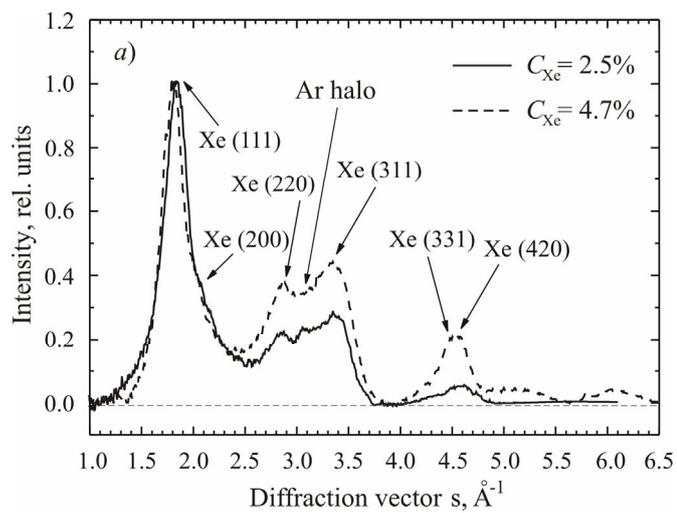

Fig. 1, *a*

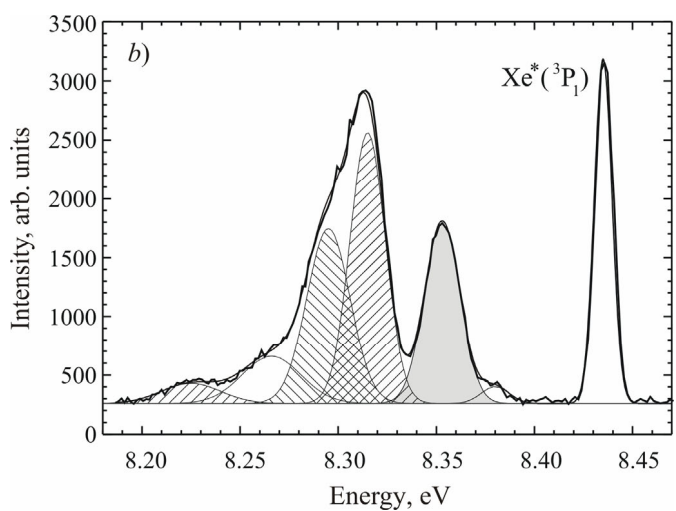

Fig. 1, *b*

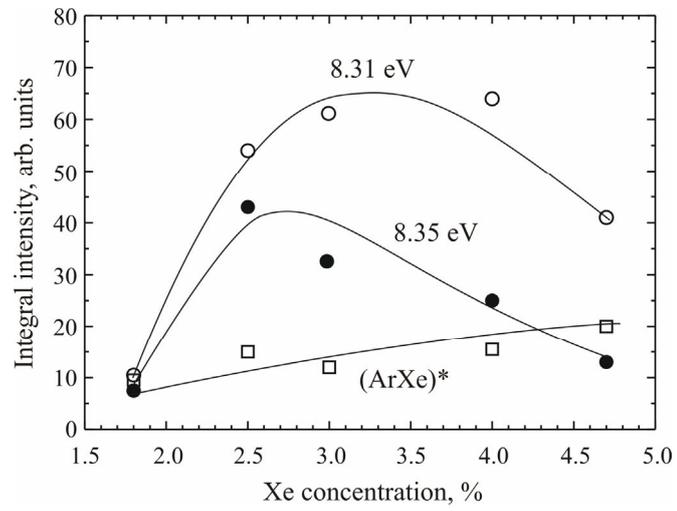

Fig. 2

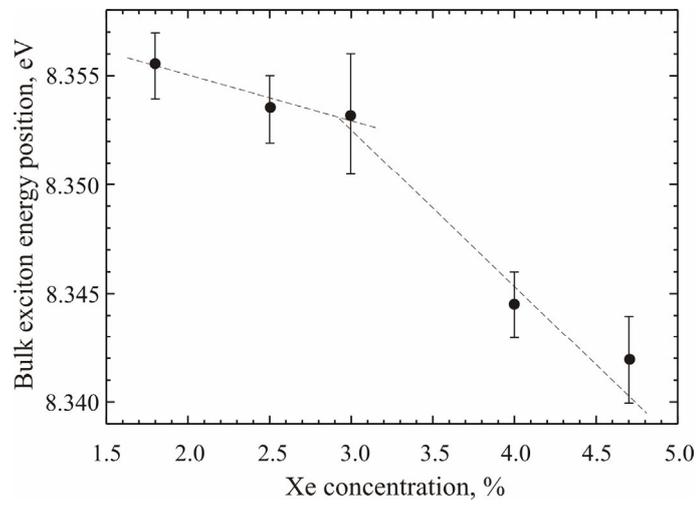

Fig. 3